\documentclass{svproc}
\usepackage{url}
\usepackage{amssymb}
\usepackage{graphicx}
\usepackage{amsmath}

\newcommand{\ba}{\begin{align}}
\newcommand{\ea}{\end{align}}%for some reason, doesn't work!
\newcommand{\beq}{\begin{equation}}
\newcommand{\eeq}{\end{equation}}

\newcommand\epsbar{{\eps_0}}
\newcommand\hdd{H^\text{DD}}
\newcommand\hssh{H_\text{SSH}}

\newcommand{\rc}{{r_{\text{C}}}}
\newcommand{\tc}{{t_{\text{C}}}}
\newcommand{\tl}{{t_{\text{L}}}}

\newcommand{\Sii}{\Si_\infty}
\newcommand{\Sisshi}{\Si_{\text{SSH},\infty}}

\newcommand{\ket}[1]{\left|#1\right\rangle}
\newcommand{\de}{\delta}
\newcommand{\eps}{\epsilon}

\newcommand{\ka}{\kappa}
\newcommand{\la}{\lambda}
\newcommand{\Si}{\Sigma}

\begin{document}
\mainmatter              % start of a contribution
\title{Qubits as edge state detectors: illustration using the SSH model}
\titlerunning{Qubits as edge state detectors}  % abbreviated title (for running head)
%$%\author{Richard MacKenzie}
%
%$%\authorrunning{Richard MacKenzie} % abbreviated author list (for running head)
%\tocauthor{Meri Zaimi, Christian Boudreault, Nou\'edyn Baspin, Hichem Eleuch, Richard MacKenzie and Michael Hilke}
%
%$%\institute{D{\'e}partement de physique, Universit{\'e} de Montr{\'e}al, Complexe des Sciences\\ C.P. 6128, succursale Centre-ville, Montr{\'e}al, QC, Canada, H3C 3J7
%$%\email{richard.mackenzie@umontreal.ca}}
\author{
Meri Zaimi\inst{1}\and
Christian Boudreault\inst{2}\and
Nou\'edyn Baspin\inst{3}\and
Hichem Eleuch\inst{4}\and
Richard MacKenzie\inst{5}\and
Michael Hilke\inst{3}
}
\institute{
Centre de Recherches Math{\'e}matiques, Universit{\'e} de Montr{\'e}al,\\ C.P. 6128, succursale Centre-ville, Montr{\'e}al, QC H3C 3J7, Canada\\
\email{meri.zaimi@umontreal.ca}\and
D\'{e}partement des sciences de la nature, Coll\`{e}ge militaire royal de Saint-Jean,\\
15 Jacques-Cartier Nord, Saint-Jean-sur-Richelieu, QC Canada, J3B 8R8\\
\email{Christian.Boudreault@cmrsj-rmcsj.ca}\and
Department of Physics, McGill University, Montr\'{e}al, QC, Canada, H3A 2T8\\
\email{nouedyn.baspin@mail.mcgill.ca}, \email{hilke@physics.mcgill.ca}\and
Department of Applied Sciences and Mathematics, College of Arts and Sciences, Abu Dhabi University, Abu Dhabi, UAE \\ and\\
%\and
Institute for Quantum Science and Engineering,
Texas A$\&$M University,\\ College Station, Texas 77843, USA\\
\email{heleuch@fulbrightmail.org}\and
D{\'e}partement de physique, Universit{\'e} de Montr{\'e}al, Complexe des Sciences,  C.P. 6128, succursale Centre-ville, Montr{\'e}al, QC, Canada, H3C 3J7\\
\email{richard.mackenzie@umontreal.ca}
}
\authorrunning{Meri Zaimi, et al.}
\tocauthor{Meri Zaimi, Christian Boudreault, Nou\'edyn Baspin, Hichem Eleuch, Richard MacKenzie (presenter) and Michael Hilke}
\maketitle              % typeset the title of the contribution

\begin{abstract}
As is well known, qubits are the fundamental building blocks of quantum computers, and more generally, of quantum information. A major challenge in the development of quantum devices arises because the information content in any quantum state is rather fragile, as no system is completely isolated from its environment. Generally, such interactions degrade the quantum state, resulting in a loss of information.\\

Topological edge states are promising in this regard because they are in ways more robust against noise and decoherence. But creating and detecting edge states can be challenging. We describe a composite system consisting of a two-level system (the qubit) interacting with a finite Su-Schrieffer-Heeger chain (a hopping model with alternating hopping parameters) attached to an infinite chain. In this model, the dynamics of the qubit changes dramatically depending on whether or not an edge state exists.  Thus, the qubit can be used to determine whether or not an edge state exists in this model.
\end{abstract}
\section{Introduction}
\label{sec-intro}
The two-level system (TLS) is the simplest nontrivial quantum system. Its simplicity notwithstanding, many important systems are TLSs. Some familiar examples are: a spin-1/2 particle (two spin states), a photon (two polarizations), a two-level atom (the two levels), a quantum dot (empty/full), two-meson systems ($K$, $\bar K$), two-flavour neutrino oscillations ($\nu_{1,2}$). Some of the above play the role of qubits, the building blocks of quantum information systems (quantum computers, teleportation, etc).

An isolated TLS, like any isolated quantum system, will evolve unitarily. This implies that pure states remain pure; assuming the two basis states are coupled, a system put in one state will oscillate back and forth between the two.

However, no system is perfectly isolated; in reality, a TLS interacts with its environment and becomes entangled with it. From the point of view of the TLS, entanglement with the environment is indistinguishable from a mixed state. We say that the pure state becomes impure, or it decoheres. In addition, in many TLSs (including the one we will study here) the interaction can permit a particle to escape from the system to the environment. In this case, from the point of view of the TLS probability is not conserved.

Decoherence and nonconservation of probability are almost always undesirable; in particular, decoherence results in a loss of information and also a loss of the potential advantage of quantum vs classical computing, quantum vs classical communication, etc. Thus, understanding (and, usually, minimizing) decoherence is critically important to the functioning of quantum devices. As an example, in \cite{eleuch2017probing} a tripartite system was studied: a TLS coupled to one end of a finite chain (or channel) whose other end was coupled to a semi-infinite chain; both chains were described by tight-binding Hamiltonians. The question addressed was: how can one reduce the decoherence of the TLS? It was found that adding noise to the channel did the trick, essentially due to Anderson localization: if excitations in the channel are localized, it becomes hard for a particle in the TLS to make its way to the far side of the channel and escape to infinity.

\begin{figure}[h!]
	\centering
	\hspace*{0cm}\includegraphics[width=0.9\textwidth]{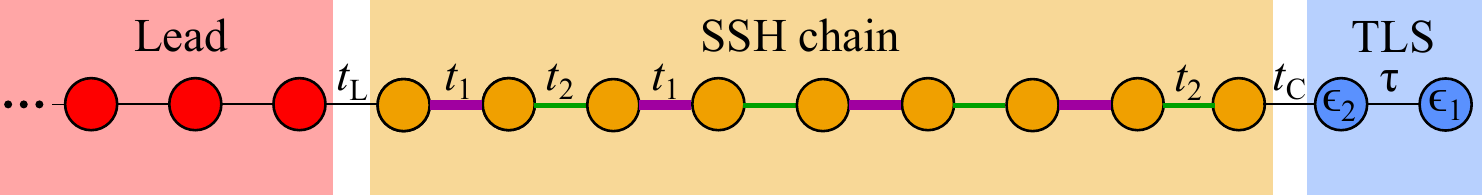}
	\caption{Tripartite system geometry. Rightmost SSH chain hopping parameter is $t_1$ or $t_2$ depending on whether number of sites $N$ is even or odd, respectively (odd case shown here).}
	\label{fig-tripartite}
\end{figure}

Here, we study a similar system with a very different goal in mind (Fig.~\ref{fig-tripartite}). The main difference is that the channel is a Su-Schrieffer-Heeger (SSH) \cite{PhysRevLett.42.1698} chain (free of disorder) described by a hopping parameter with alternating hopping strengths. Such chains have topological edge states (for a review, see \cite{EE1}), and rather than trying to minimize the decoherence of the TLS, we imagine measuring its decoherence rate to determine whether the system to which it is attached has edge states. As we will see, the presence of edge states greatly increases the decoherence rate.

%%%
\section{Two-level system: a rapid review}
\label{sec-tls}
We review the isolated TLS, mostly to establish notation to be used in what follows. The TLS Hamiltonian is
\beq
\hdd=\left( \begin{matrix}
	\eps_2 & \tau\\
	\tau & \eps_1
\end{matrix} \right)
\equiv
\left( \begin{matrix}
	\eps_0-\de_0/2 & \tau\\
	\tau & \eps_0+\de/2
\end{matrix} \right).
\label{eq-H_DD}
\eeq
The energies are
$\la_\pm = \frac{1}{2}(\eps_1+\eps_2 \pm\de) = \epsbar \pm \frac{\de}{2}$,
where $\de = \sqrt{(\eps_1-\eps_2)^2+4\tau^2}$.

The energy-dependent Green's function is defined by $G^\text{DD}(E) = (E-\hdd)^{-1}$; its Fourier transform gives the time-dependent Green's function, which is a sum of oscillatory terms with frequencies given by the energies; for instance,\beq
\label{G12t-isolated}
G^\text{DD}_{12}(t)=-\frac{2\pi i\tau}{\de}\left(e^{-i\la_+t}-e^{-i\la_-t}\right).
\eeq
When we couple the TLS to the rest of the system, it will decohere; this will be seen in the Green's function, which will exhibit decaying behavior \cite{eleuch2017probing}.

%%%
\section{Su-Schrieffer-Heeger model and edge states}
\label{sec-ssh}
The SSH Hamiltonian \cite{PhysRevLett.42.1698}, proposed in the context of the polymer polyacetylene for reasons we will not go into here, is
\beq
\hssh=\left(\begin{matrix}
0 ~~& t_1 \\
t_1 ~~& 0 ~~& t_2 ~~&\\
~~& t_2 ~~& 0 ~~& t_1 ~~& \\
~~& ~~& t_1 ~~& 0 ~~& \ddots ~~&  \\
~~& ~~& ~~& \ddots ~~& \ddots ~~& t \\
~~& ~~& ~~& ~~& t ~~& 0 \\
\end{matrix}\right),
\label{eq-H_SSH}
\eeq
where $t=t_1$ or $t_2$ for $N$ even or odd, respectively. We will assume $t_1,t_2>0$ for simplicity, and for now we assume $N$ is even and write $N=2M$. Much of what follows is known \cite{delplace2011zak,ref-asboth,PhysRevB.97.195439,us}; we repeat it to establish notation and to focus on results to be used below.

To solve the Schroedinger equation, translational invariance (by two sites) suggest the following ansatz:
\beq
\ket{\psi}=\sum_{n=0}^{M-1} \left( A\ket{2n+1}+B\ket{2n+2} \right) e^{in2k}.
\label{eq-psi1}
\eeq
We can take $k$ between $\pm\pi/2$ since $k\to k+\pi$ has no effect on $\ket\psi$.
The middle components of the Schroedinger equation (all but the first and last) determine the dispersion relation and also the ratio $A/B$. The former is
\beq
E^2 = t_1^2 + t_2^2 + 2 \,t_1 t_2 \cos 2k.
\label{eq-Esq}
\eeq
Assuming $k$ is real, $(t_1-t_2)^2<E^2<(t_1+t_2)^2$ so there are two energy bands. For any allowed energy, \ref{eq-Esq} has two equal and opposite solutions for $\pm k$ where we assume $k>0$. Thus the general solution to the middle equations is a linear combination of the solutions for $\pm k$.

The edge components of the Schroedinger equation (the first and last) determine the ratio of these two solutions, and also the energy eigenvalues. The latter are given by the solutions of the following equation for $k$, where $r=t_1/t_2$ and we have written $\text{s}_j = \sin(j k)$.
\beq
r \, \text{s}_{N+2} + \text{s}_N = 0.
\label{eq-k}
\eeq
where $r=t_1/t_2$ and we have written $\text{s}_j = \sin(j k)$.

\begin{figure}[h!]
	\centering
	\hspace*{0cm}\includegraphics[width=0.4\textwidth]{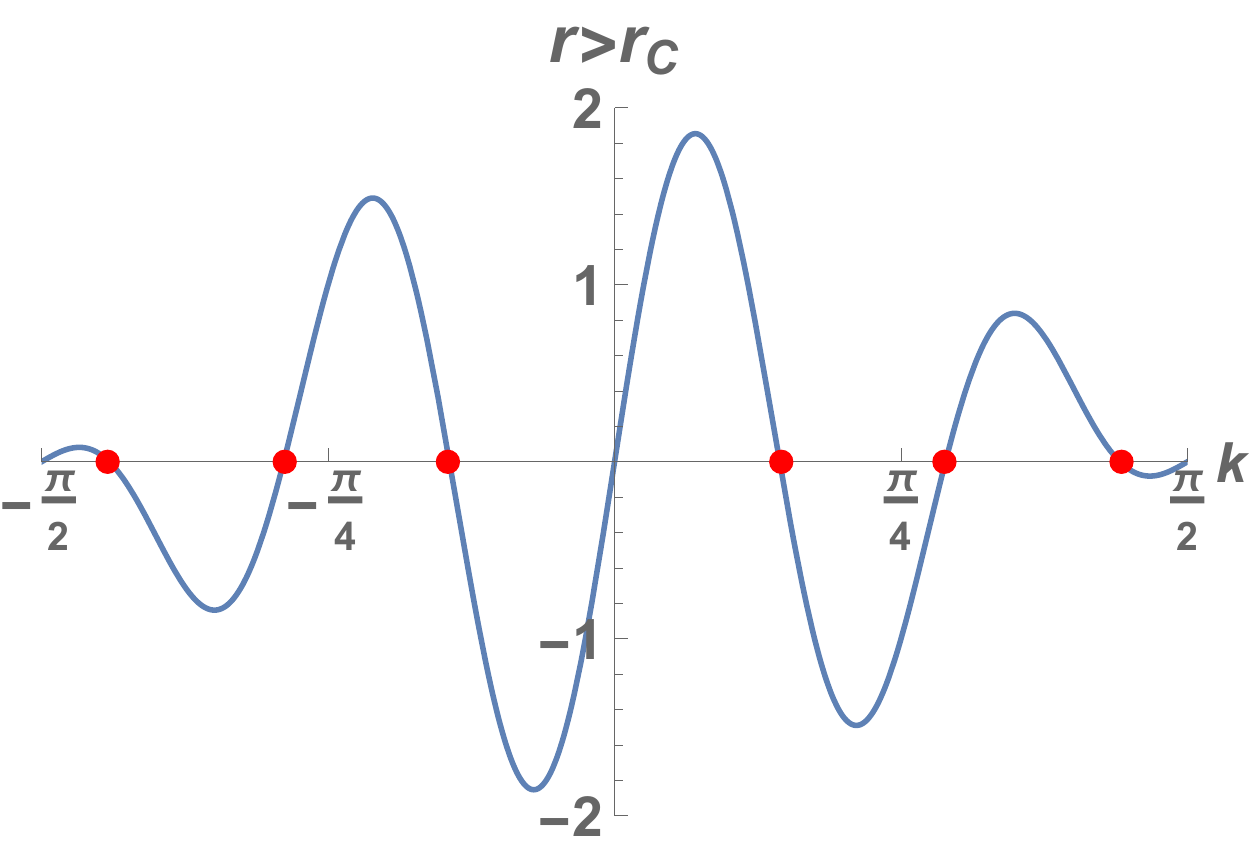}~~
	\includegraphics[width=0.4\textwidth]{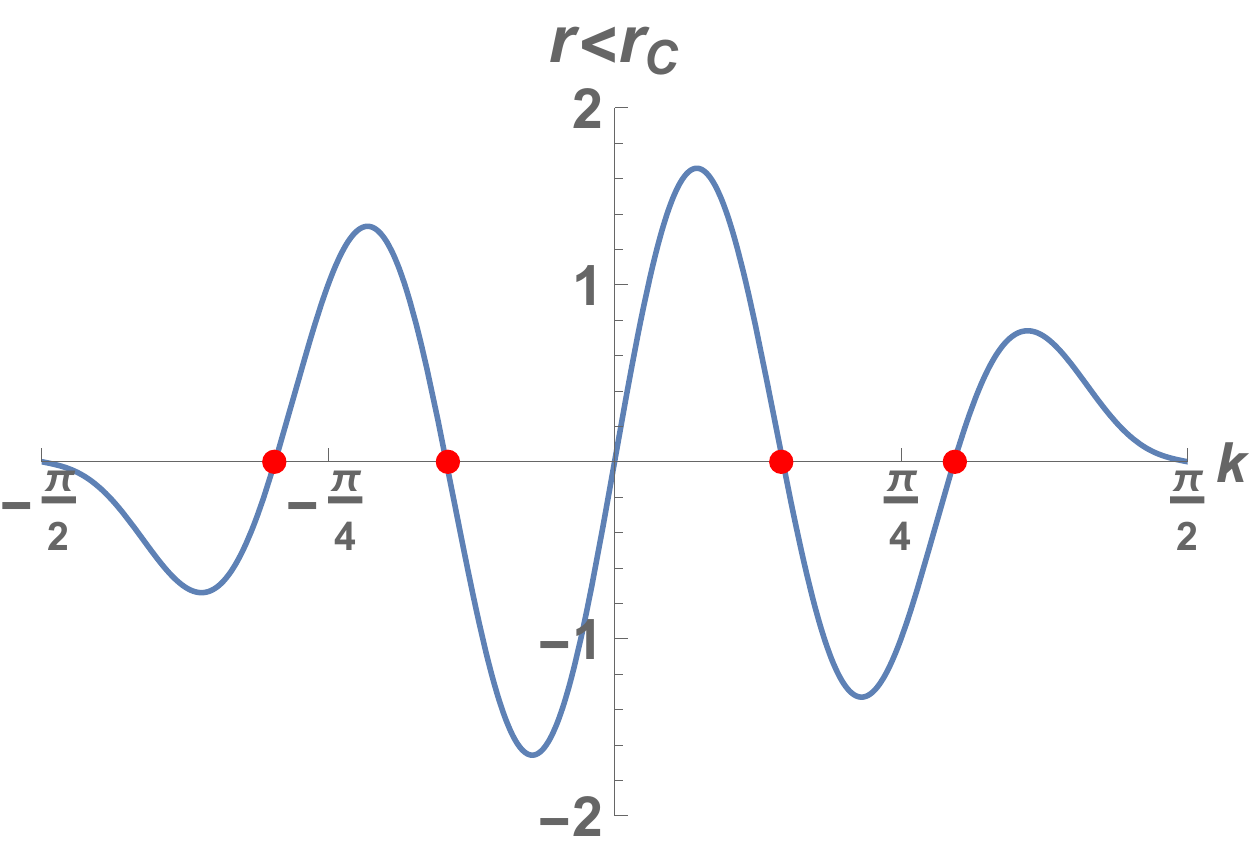}
	\caption{Graphical solution of \eqref{eq-k} for $N=6$ ($\rc=0.75$). Left panel: $r=0.9$; six solutions. Right panel: $r=0.7$; four solutions. (Note that $k=0,\pm\pi/2$, although solutions of \eqref{eq-k}, do not correspond to solutions to the SE.)}
	\label{fig-graphical}
\end{figure}

This equation cannot be solved analytically; however, numerically or graphically (see Fig.~\ref{fig-graphical}) we find that there are $N$ real solutions, as required, for $r>\rc$ whereas there are two fewer real solutions for $r<\rc$, where \cite{delplace2011zak}
\beq
\rc \equiv \frac{N}{N+2}.
\eeq

\begin{figure}[h!]
	\centering
	\hspace*{0cm}\includegraphics[width=0.7\textwidth]{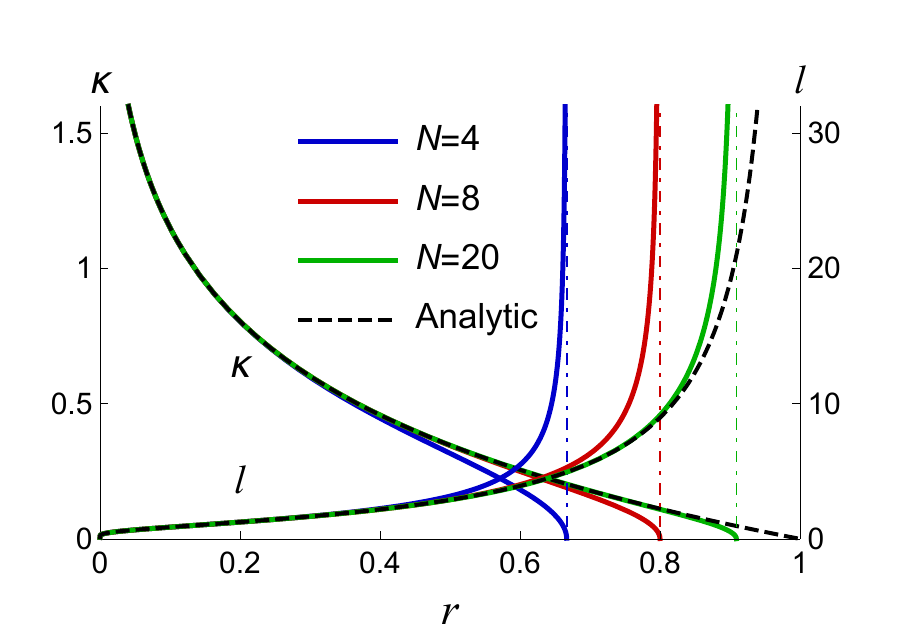}
	\caption{Decay rate $\ka$ and decay length $l$ of edge states for various values of $N$. Also displayed is an analytic solution to \eqref{eq-kappa} for $N\to\infty$.}
	\label{fig-kappa}
\end{figure}

Thus, for $r<\rc$ there are two missing solutions. These turn out to be solutions of complex wave number, $k=\pi/2 \pm i\ka$, where $\ka$ is the positive solution of
\beq
\frac{\sinh(N\ka)}{\sinh((N+2)\ka)} = r.
\label{eq-kappa}
\eeq
the solution of which is displayed in Fig.~\ref{fig-kappa} for various values of $N$. These states, having complex $k$, are exponentially confined to the edges of the system: they are edge states. Also displayed is $l=1/\ka$, the penetration length of the edge states. As $r\to\rc$ from below, we see that the length scale goes to infinity; the ``edginess" of the edge states becomes irrelevant if $l\gg N$.

\begin{figure}[h!]
	\centering
	\hspace*{0cm}\includegraphics[width=0.48\textwidth]{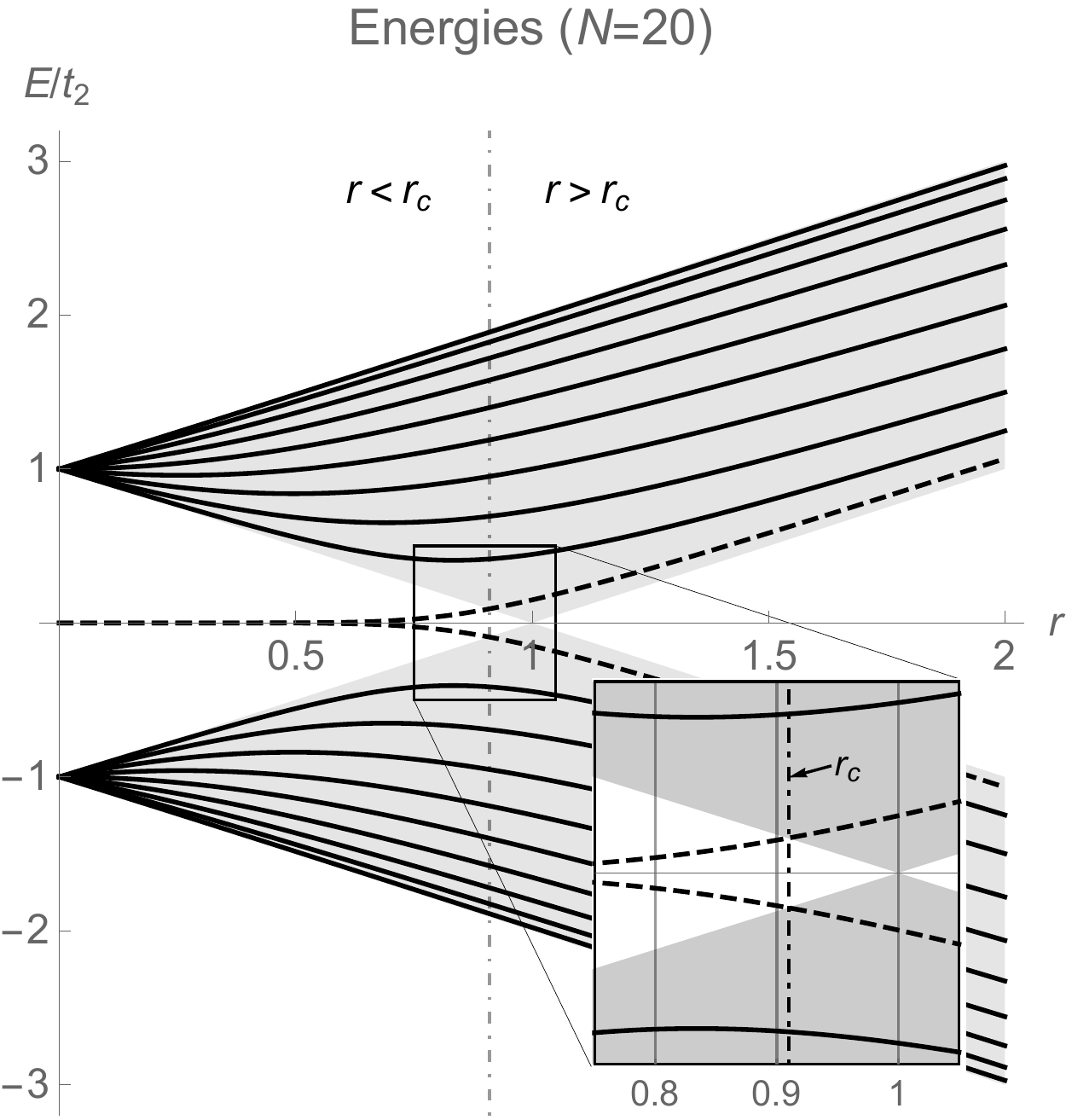}~~
	\includegraphics[width=0.48\textwidth]{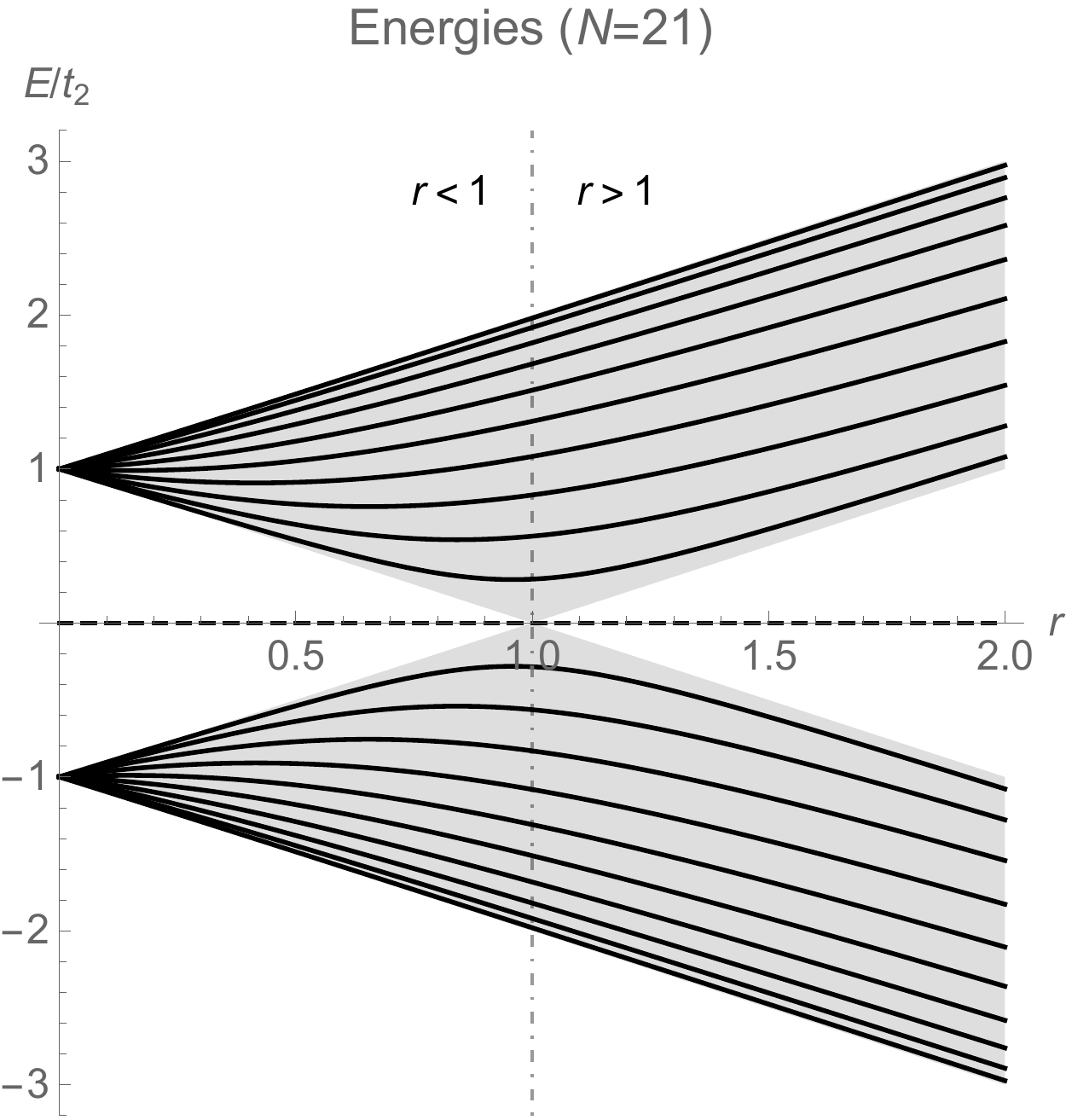}
	\caption{Energy spectra for two values of $N$ as a function of $r$. The shaded regions are the bands for $N=\infty$. Dotted lines outside the bands are edge states. The inset on the left focuses on the transition between an edge state (to the left of the vertical broken line) and a non-edge state (to the right, in the shaded region).}
	\label{fig-spectra}
\end{figure}

We conclude with a brief discussion of the case $N$ odd, which is in fact much simpler. It is easy to show that no matter the value of $r$, there is always exactly one zero-energy edge state (the remainder of the spectrum being symmetric). This state is confined to the left (right) edge for $r<1$ ($r>1$) with decay length $l=1/|\log r|$. Fig.~\ref{fig-spectra} displays the spectra for $N=20$ and 21.

%%%
\section{Tripartite system: TLS-SSH-chain}
\label{sec-tri}
We now study the tripartite system displayed in Fig.~\ref{fig-tripartite}. Although it is an infinite-dimensional system, the effects of the SSH chain and semi-infinite chain on the TLS can be nicely incorporated into a $2\times2$ effective Hamiltonian for the TLS; these effects are simply given by a term added to the $(1,1)$ component of the Hamiltonian \cite{us,datta2005quantum}:
\beq
\eps_2\to\eps_2+\Sisshi \equiv \eps_2'.
\label{eq-epsprime}
\eeq
Here $\Sisshi$ is proportional to the surface Green's function of the combined SSH chain and semi-infinite chain. This can be calculated analytically, although it is fairly nasty. The result is \cite{us}
\beq
\Sisshi = 
\begin{cases}
\displaystyle
\tc^2 \frac{Et_2{s}_N-\Sii(t_1{s}_{N-2}+t_2{s}_N)}{t_2^2(t_1{s}_{N+2}+t_2{s}_N)-Et_2\Sii{s}_N} & (N\text{ even}) \\ \\
\displaystyle
\tc^2\frac{t_2(t_2{s}_{N-1}+t_1{s}_{N+1})-E\Sii{s}_{N-1}}{t_1t_2E{s}_{N+1}-t_1\Sii(t_2{s}_{N+1}+t_1{s}_{N-1})} & (N\text{ odd})
\end{cases}
\label{sig_Analytical}
\eeq
where
\beq
\Sii = \frac{\tl^2}{2}\left(E-i\sqrt{4-E^2}\right).
\label{eq-Siginfp}
\eeq
Note that $\eps_2'$ is complex, so the effective Hamiltonian is no longer Hermitian. This is related to the open nature of the TLS: being non-Hermitian, time evolution preserves neither probability nor purity, reflecting the fact that the particle can escape to its environment, and that the TLS and environment become entangled.

Defining $\la_\pm'$ and $\de'$ as the quantities defined in Section \ref{sec-tls} with the substitution \eqref{eq-epsprime}, we can use these substitutions in the definition of $G^\text{DD}(E)$ given earlier to get the new energy-dependent Green's function, $G^\text{DD}_{\text{SSH},\infty}(E)$. It is tempting to suppose that these substitutions also work for the time-dependent Green's function. This is not quite correct, since $\la_\pm'$ depend in a highly nontrivial way on $E$ so the Fourier transform cannot be evaluated exactly. An analytical approximation which is justified in the weak-coupling limit ($\tc\ll1$) \cite{eleuch2017probing} indicates that to a good approximation the new (complex) frequencies $\la_\pm'$ can be evaluated at the old frequencies: the time-dependent Green's function has, according to this approximation, frequencies $\la_\pm'(\la_\pm)$.
According to this analytic approximation, the decay rates are given by the imaginary part of the frequencies, and we conclude that the decoherence time $\tau_\phi$ is given by
\beq
\left(\tau_\phi\right)^{-1} \approx \min \left( -\frac{1}{2}\Im\left\{\Sisshi(\la_\pm)
\pm\de'(\la_\pm)\right\} \right).
\label{eq-coupletolead2}
\eeq

\begin{figure}[h!]
	\centering
	\hspace*{0cm}\includegraphics[width=0.48\textwidth]{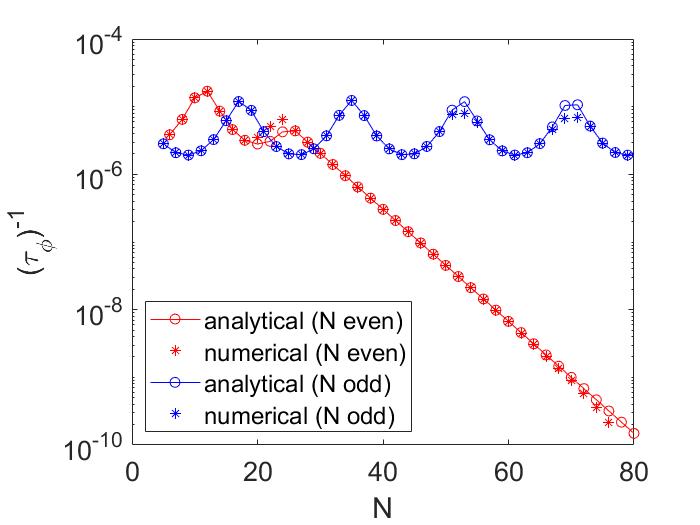}~~
	\includegraphics[width=0.48\textwidth]{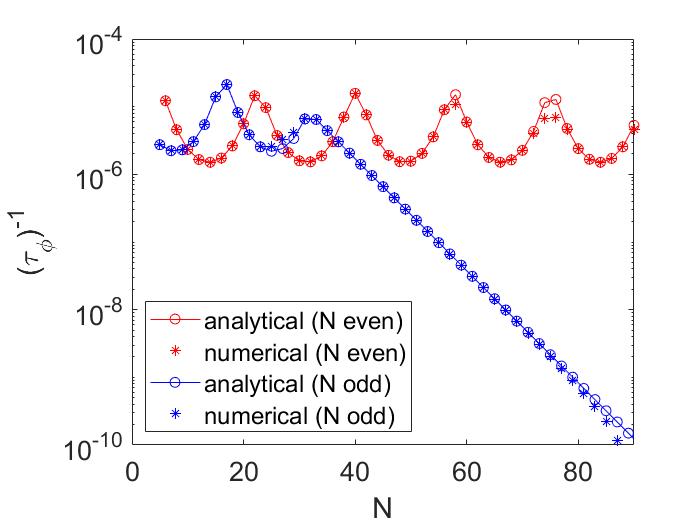}
	\caption{Decoherence rate as a function of $N$ for $t_1=1/t_2=1.1$ (left), $t_2=1/t_1=1.1$ (right). For both figures, $(\eps_1,\eps_2,\tau,\tc,\tl)=(.4022,.0022,.03,.035,.65)$. The values for $\eps_{1,2}$ were chosen so that the isolated TLS has a zero eigenvalue, corresponding exactly ($N$ odd) and approximately ($N$ even) to the edge state energy.}
	\label{fig-decay}
\end{figure}

This analytical approximation can be justified post hoc by comparing \eqref{eq-coupletolead2} with a numerical evaluation of the decoherence rate. Both are displayed in Fig.~\ref{fig-decay}.
The figure, which encapsulates our main result and merits some discussion. The graph on the left corresponds to $r=1.21$, for which there are no edge states if $N$ is even, while there is a right edge state if $N$ is odd. We see dramatically different behavior depending on the parity of the chain. If $N$ is even (red, lower curve), there are no edge states, so decoherence is suppressed: for $N$ sufficiently large, the rate decreases exponentially with $N$. If $N$ is odd (blue, upper curve), the right-hand edge state couples strongly to the TLS and the decoherence remains large as $N$ increases.

The graph on the righ corresponds to $r\sim0.83$, for which there are two edge states if $N$ is greater than 10 and even, while there is a left edge state if $N$ is odd. Again, the behavior is dramatically different for even vs. odd parity. If $N$ is even (red, upper curve), the presence of edge states maintains a high decoherence rate as $N$ increases. If $N$ is odd (blue, lower curve), the absence of an edge state on the TLS side of the SSH chain gives rise to exponential decoherence suppression as $N$ increases.

%%%
\section{Conclusions}
\label{sec-conclusion}
The interaction between a TLS and its environment can have a strong effect on the dynamics of the TLS. Here, we argued that coupling to one end of an SSH chain (which is coupled at the other end to an undimerized infinite chain) can have a very strong effect on the decoherence of the TLS. The effect is dramatically different depending on whether there is or is not an edge state at the TLS end of the SSH chain: an edge state causes decoherence to remain high independent of chain length, wereas in the absence of an edge state decoherence decreases exponentially with chain length. This suggests using a TLS as a sort of edge state detector.

\section*{Acknowledgments}
This work was supported in part by the Natural Science and Engineering Research Council of Canada, by the Fonds de Recherche Nature et Technologies du Qu{\'e}bec via the INTRIQ strategic cluster grant, and by the Perimeter Institute for Theoretical Physics. Research at Perimeter Institute is supported by the Government of Canada through the Department of Innovation, Science and Economic Development and by the Province of Ontario through the Ministry of Research, Innovation and Science.


\end{document}